\documentclass[12pt]{iopart}
\usepackage{epsf} 
\begin{document}

\title[Gravitational wave burst vetoes in the LIGO S2 and S3 data analyses]{Gravitational wave burst vetoes in the LIGO S2 and S3 data analyses}

\author{Alessandra Di Credico\footnote[1]{on leave from Laboratori Nazionali del Gran Sasso (INFN, Italy)}(for the LIGO Scientific Collaboration)}

\address{Physics Department, Syracuse University, Syracuse NY 13244, USA}

\begin{abstract}

The LIGO detectors collected about 4 months of data in 2003--2004
during two science runs, S2 and S3. Several environmental and auxiliary
channels that monitor the instruments' physical environment and overall
interferometric operation were analyzed in order
to establish the quality of the data as well as the
presence of transients of non-astrophysical origin.
This analysis allowed better
understanding of the noise character of the instruments and
the establishment of correlations between transients in these channels
and the one recording the gravitational wave strain. In this way
vetoes for spurious burst events were identified.
We present the methodology we followed in this analysis and the
results from the S2 and S3 veto analysis within the context of the
search for gravitational wave bursts.

\end{abstract}



%\maketitle

\section{Introduction}

The Laser Interferometer Gravitational wave Observatory (LIGO)
collected data during the years 2003 and 2004 in two distinct science runs: 
S2 (February 14,~2003 - April 14,~2003) and S3 (October 31,~2003 - 
January 9,~2004).
Using this data a search for gravitational wave bursts was
conducted and, for the S2 run, results are to be published soon~\cite{paperS2}.
The search for gravitational wave bursts represents a great challenge for the 
data analysis, since it looks for signals that are very poorly (if at all)
modelled and of unknown (although very short) duration.
For most of the LIGO searches for bursts so far, the 
frequency range has covered the best sensitivity range for the LIGO 
detectors (100--1100Hz). 

Several methods exist for the identification of candidate burst events
in LIGO.
In S2 and S3 the search result was obtained by using a wavelet-based 
algorithm called WaveBurst~\cite{WaveBurst1,WaveBurst2}.
WaveBurst identifies clusters of excess power in the time-frequency
plane once the signal is decomposed in the wavelet domain.
This Event Trigger Generator (ETG) processes gravitational wave data from two
interferometers at a time.
Following this event selection by WaveBurst we checked
the consistency of the waveforms of the candidate
events.
This was done using the $r$-statistic~\cite{rstat},
a time-domain cross-correlation method sensitive to the coherent part
of the candidate signals.

\section{Data Quality}
\label{DataQ}

A key first step in the search for gravitational wave bursts
is establishing the overall quality of the data produced by the
interferometers. 

In both the S2 and S3 searches for bursts with LIGO, and given
the sensitivity of the instruments, we limited our search in the 
100--1100~Hz frequency regime.
Within this frequency range there are still sources of non-gaussian noise, 
which especially affect the quality of the data when it is to be searched 
for burst signals.

Several algorithms running in real-time with the data acquisition
(``monitors'') keep track of the statistical properties of the noise in
the detectors.
One such algorithm is the band-limited root-mean-square, {\it{BLRMS}}, 
monitor of a channel's amplitude power within a frequency band.
Several frequency bands are monitored and trends of the in-band power 
are continuously recorded.
Data quality flags based on the {\it{BLRMS}} monitors executed on
the gravitational wave channel as well as many other channels have been
traditionally used to signal noisy or overall problematic data
taking~\cite{paperS1}.
During S2, excursions of the {\it{BLRMS}} monitor running on the
gravitational wave channel in the 4km Hanford interferometer (H1)
were seen to be resulting from instabilities in the servo loops.
A data quality cut based on this was derived and implemented in the
S2 search for bursts.
The cut selected instances when the {\it{BLRMS}} exceeded a threshold of 
0.0002~ct$^2$/Hz in power
over a five minute interval in H1 in the 200--400~Hz band.
This cut resulted in a loss of live-time of $\sim$0.4\%.
No similar {\it{BLRMS}} excursions of the gravitational wave channels
persisting over five minutes (or more) were seen during S3.

Calibration lines (fixed amplitude harmonic excitations)
are continuously injected into the end test masses of the
LIGO instruments during the science runs.
This is in order to enable us to monitor the detector's sensitivity 
and frequency response throughout a run.
We require the calibration lines to be always present and
their amplitudes strong enough so that the calibration of a 
gravitational wave strain measurement is reliable.
During S2, this requirement reduced the triple coincidence live-time by 
about 2\% and 2.4\% in S3.

There are occasional losses of synchronization, timing issues, or overall data
acquisition related problems that may result in loss of data.
These affected overall small fractions of the instruments' triple coincidence 
live-time, at the level of 0.3\% in S2 and about 5\% in S3. 

The reduction of the noise floor of the LIGO instruments in the
last two years has revealed previously invisible couplings
with their physical environment.
For this reason our veto search requires ongoing investigation. 

One such coupling is of seismic origin: seismic activity is present 
in all detector locations but its effects on the gravitational wave channel
were mostly noticed in the Hanford site. 
Its origin has been connected to human activity like gravel 
trucks driving by the Hanford Observatory or natural phenomena like 
strong winds or local earthquakes.
During S3 we observed elevated interferometer noise during long stretches of
high seismic activity as identified by the {\it{BLRMS}} monitor. 
Looking at the output of a seismometer located in the corner building of
the Hanford site, we set a flag when the seismic noise was too high.
A cut derived from this flag reduced the S3 data by 1.3\%.

\begin{figure}[ht]
\begin{center}
\epsfxsize=5.0in
\mbox{\epsfbox{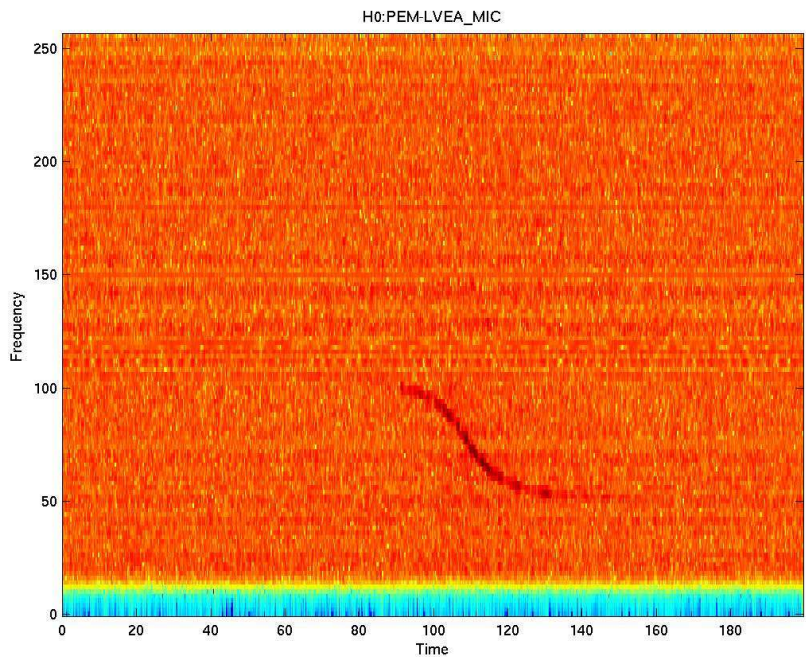}}
\caption{\label{fig:acoustic}Spectrogram of the acoustic noise observed 
with a microphone located in the proximity of the Hanford detector, at 
the time an airplane was flying over the site} 
\end{center}
\end{figure}

Acoustic coupling of the LIGO instruments to their environment has been
observed since the early engineering runs through the manifestation
in the interferometric data of overflying airplanes or strong winds.
Several microphones are located in the vicinity of the test masses,
laser and output photodyodes.
Elevated acoustic ambient noise is identified via a {\it{BLRMS}} 
excursion of power in these microphonic channels.
In the case of overflying airplanes, the microphones record an additional
typical Doppler shifted acoustic signature in the 50--100~Hz frequency range,
with a duration of tens of seconds.
A typical example of the signal recorded by one of the microphones
in the vicinity of the Hanford instruments when an airplane is flying over 
the site is displayed in figure~\ref{fig:acoustic}.

The use of the ambient acoustic noise as a data quality cut was introduced
in our S3 burst analysis.
This is after the S2 search for bursts yielded a single candidate
event that was found to be of acoustic origin in the Hanford
detectors~\cite{paperS2} and was found to correlate perfectly
and as expected from direct measurements
with elevated acoustic power in the microphones.
Between S2 and S3 significant work took place in order to reduce the
acoustic coupling of the instruments.
Although this resulted to 2 or 3 orders of magnitude reduction of such
coupling, it is still non-zero and for this the acoustic data quality cut
is envisioned to be used in the search for bursts in S3 and beyond.
For the S3 analysis a flag of acoustic noise was set whenever the band limited 
RMS, for the frequency band 62--100~Hz, exceeded a threshold value 
set indipendently at the Hanford and Livingston sites. 
This resulted in a live-time loss of less than 1\%.

\section{Event-by-event vetoes}
\label{EventVeto}

After excising the data that fail the data quality cuts described above 
we analyze the remaining gravitational wave burst data with
the burst-finding algorithms.
The outcome of this search is candidate events that reflect the time-frequency
and waveform consistency among signals in the three LIGO instruments.
Since our first search for gravitational wave bursts~\cite{paperS1}, we have
allowed in our search pipeline the incorporation of event-by-event
vetoes.
These vetoes are derived from environmental and auxiliary channels
and indicate disturbances of non-astrophysical origin that couple
to the gravitational wave strain channel.

Our analysis starts with a subset of the data, 
about 10\% of the total, that is called ``playground''.
The concept of playground was introduced in the first LIGO analyses
in order to look at a representative fraction of the data, define
the analysis parameters there and then apply them to the rest of the data
without introducing a bias. 

Most of the results that we will show here are derived from triggers
generated by the {\it{glitchMon}} and {\it{KleineWelle}} methods.
{\it{glitchMon}}~\cite{glitchMon} is a time-domain transient-finding 
algorithm developed within the Data Monitoring Tool (DMT) environment.
It allows the selection of times when a channel's amplitude
exceeds a user-defined threshold, either in absolute ADC counts
or in units of amplitude rms.
{\it{KleineWelle}}~\cite{KWgwdaw8} is also developed within the
Data Monitoring Tool (DMT) environment.
It is a time-frequency method offering the multi-resolution approach
of the wavelet decomposition.
Within {\it{KleineWelle}} the time series
is first whitened using a linear predictor error filter.
Then the time-frequency decomposition is obtained through the Haar
wavelet transform.
The squared wavelet coefficients, normalized to the scale's rms,
provide an estimate of the energy associated with a certain time-frequency
pixel.
A clustering mechanism is then invoked, in order to increase the
sensitivity to signals  with less than optimal shapes in the time-frequency
plane, and a total normalized cluster energy, E$_C$ is computed.
This quantity, like the previous single pixel normalized energy,
is $\chi^2$  distributed (for gaussian white noise).
The significance {\it S} of this cluster of pixels is then defined
using the standard definition for a $\chi^2$ distribution with {\it N} 
degrees of freedom:

\begin{equation}
S = - ln \int^\infty_{E_C}{\chi^2_N (E) dE}
\end{equation}

The significance is a function of the cluster energy and the number of pixels
in the cluster, {\it N}, and is the parameter used in the threshold choice 
for the KleineWelle transient production.

The ultimate goal of the veto data analysis is to establish a correlation
between spurious triggers in the gravitational wave channel and triggers 
in auxiliary channels, above the rate of accidental coincidences.
For our purposes the coincidence is defined by the overlap of the 
gravitational wave trigger (thought of as a window of a given start time 
and stop time) with a trigger (thought of in the same way) from an 
auxiliary channel. 

The central notions in analysing an auxiliary channel
as a potential veto are the following:
\begin{itemize}
\item {\it{efficiency:}} this is the percentage of gravitational wave events
which are vetoed by the auxiliary channel triggers,
\item {\it{use percentage:}} this is the percentage
 of auxiliary channel trigger that veto a gravitational wave event,
\item {\it{dead-time:}} this is the percentage of the instruments' live-time
reflecting the total duration of the auxiliary channel triggers, 
\item {\it{accidental coincidences:}} this is the number of overlaps
between the gravitational wave and auxiliary channel triggers that
may occur from pure chance,
\item {\it{safety:}} this is the assurance that the coupling of the auxiliary
channel to a genuine gravitational wave signal is null, or below a threshold of
detectability.
\end{itemize}

\begin{figure}[ht]
\begin{center}
\epsfxsize=3.0in
\mbox{\epsfbox{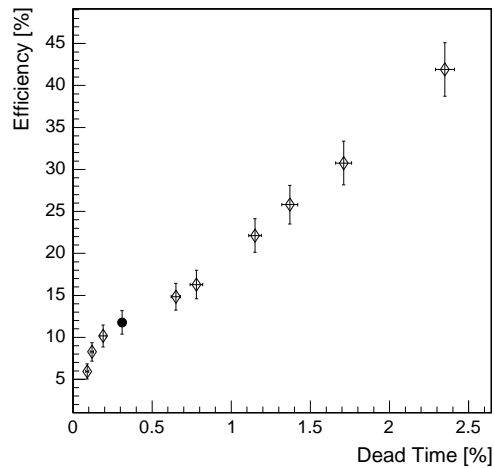}}
\caption{\label{fig:PRCCTRLeffvsdt}Efficiency versus dead-time study for 
H2:LSC-PRC\_CTRL. The empty diamonds refer to different choices of threshold
for the transient significance: from right to left: 20,30,40,50,80,100,300,
400,500. The full circle refers to the threshold chosen in the S3 
veto analysis}
\end{center}
\end{figure}

Our search for vetoes includes the calculation of the above quantities
for multiple auxiliary channels as well as for varying thresholds
for both the auxiliary channel and the gravitational wave channel.
Several plots and derived quantities starting with the above can guide
us in making veto choices in a search.
A channel with high efficiency and low dead-time is clearly desirable.
In figure~\ref{fig:PRCCTRLeffvsdt} we show a typical such plot for 
H2:LSC-PRC\_CTRL
(see below in subsection~\ref{VetoS3} for a description of this channel).
The curve traces
a total of 9 different thresholds of the auxiliary channel.
 
\begin{figure}[ht]
\begin{center}
\epsfxsize=3.7in
\mbox{\epsfbox{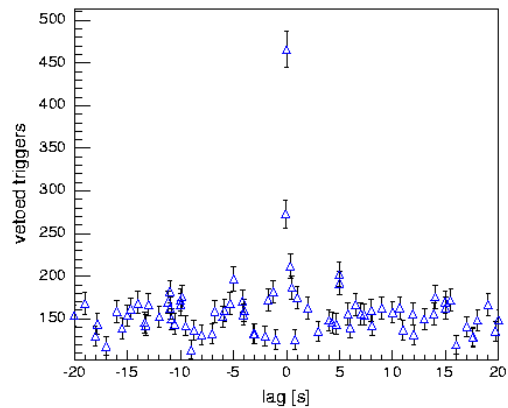}}
\caption{\label{fig:PRCCTRLlagplot}Real coincidences appear at lag 
(time-shift) = 0 sec in this plot for H2:LSC-PRC\_CTRL. Random coincidences 
are seen when the lag is either positive or negative}
\end{center}
\end{figure}

Another consideration in choosing a veto channel is the use percentage
of its triggers: the higher being the better.
The accidental coincidences between the gravitational wave and auxiliary
channel triggers are evaluated by introducing artificial time shifts
between the two trigger lists.
This is shown in figure~\ref{fig:PRCCTRLlagplot} for the case of 
H2:LSC-PRC\_CTRL.

The above quantities are generally interconnected and a decision
on the choice of a veto can not be readily automated.
So far we have taken a conservative approach requiring potential
vetoes to result in minimal dead-time (of order of few percent)
with an efficiency of 10\% or more.
While this procedure is reasonable for an upper-limit search,
a more inclusive approach in accepting vetoes is being examined
in a detection-oriented search.
Part of this direction includes the study of the efficiency of
vetoes to reject outliers of single-interferometer gravitational
wave triggers as well as basing choices and thresholds of vetoes
on direct measurements of their coupling to the gravitational wave
channel.

In order to assess the safety of potential
veto channels, we have used a set of signals injected directly into
the LIGO interferometers 
This allows us to measure the coupling of a veto channel to the gravitational 
wave channel or to establish an upper-limit to the coupling.
This is essential for making sure that such veto does not reject an 
astrophysical event.

\subsection{Vetoes in S2}
\label{VetoS2}

The veto analysis for the S2 data considered several auxiliary channels,
including those which track possible environmental signals.
Most of these channels did not make good veto candidates as they
showed efficiencies typically below 10\%, accompanied
by dead-times of around 5\%.
The most interesting candidate veto channel was found in the
Livingston inteferometer, L1:LSC-AS\_DC. This channel reads 
the DC level of the light output in the antisymmetric port of the instrument, 
and was observed to correlate with the gravitational wave
channel through a non-linear coupling with interferometer alignment
fluctuations.
The efficiency level for this channel reached 15\% with a dead-time of 5\%,
both of marginal interest for accepting it as a veto.
In a conservative approach we decided not to use it in the S2 search.

\subsection{Vetoes in S3}
\label{VetoS3}

The study of vetoes in the S3 playground yielded
a few auxiliary channels which presented an interesting correlation
with the gravitational wave triggers.
Among these were 
LSC-AS\_I for the 4km Hanford interferometer
(H1) and LSC-PRC\_CTRL for the 2km 
Hanford one (H2). Both these channels show a veto efficiency above 10\% 
and a dead-time of less than 1\%. 

The LSC-AS\_I channel is tightly connected to the gravitational wave 
channel (LSC-AS\_Q) as both of these signals are extracted from the 
antisymmetric port signal. LSC-AS\_I records the  ``In phase'' part of 
the signal while LSC-AS\_Q records the ``Quadrature phase'', the part 
of the signal 
directly related to the gravitational wave strain. Given the common 
source, small mixing between the two channels is possible, and the safety 
of the LSC-AS\_I channel must be established with special care.

The safety study on LSC-AS\_I has been conducted also by the inspiral 
search working group and has brought similar results: this auxiliary 
channel can be considered safe only when a threshold is imposed on the 
significance of the triggers and the ratio between the LSC-AS\_I and 
the gravitational wave trigger significances is above 0.5~\cite{InspVeto}. 

In the burst veto search, when these safety conditions are applied, the 
efficiency of LSC-AS\_I reaches levels too low to be acceptable.

Another auxiliary channel that has shown interesting efficiency in 
vetoing the gravitational wave channel is H2:LSC-PRC\_CTRL.
This channel is related to the power recycling cavity and monitors 
the feedback loop that keeps it in resonance. 
When a high-pass filter at 70Hz is applied to the signal, a good 
correlation between the transients observed in this channel and those 
seen in the gravitational wave channel has been observed. The highest 
veto efficiency reached with this auxiliary channel is of about 40\%.

In Figure~\ref{fig:PRCCTRLeffvsdt}, values of efficiency versus dead-time for 
different significance thresholds are shown. The dead-time, 
computed considering only the duration of the lowest threshold transients, 
is about 2.4\%.
  
\begin{figure}[ht]
\begin{center}
\epsfxsize=3.0in
\mbox{\epsfbox{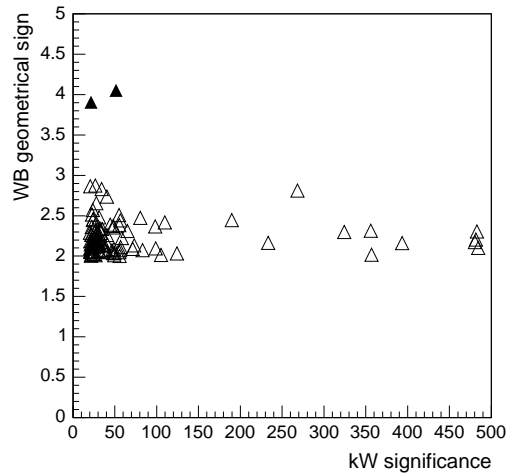}}
\caption{\label{fig:PRCCTRLsafety}Veto safety study for H2:LSC-PRC\_CTRL. 
Coincidences with reconstructed hardware injection signals are plotted as 
full triangles. Coincidences between the same auxiliary channel transients 
and the gravitational wave channel triggers are plotted as empty triangles.}
\end{center}
\end{figure}

The choice of the threshold for the veto channel is determined mainly by the 
dead-time and by the safety of the veto. Figure~\ref{fig:PRCCTRLsafety} 
shows the result of the safety study for H2:LSC-PRC\_CTRL. A few burst 
hardware injections signals were found in coincidence with low significance 
auxiliary channels transients. A conservative threshold 
of 200 has thus been chosen to guarantee the safety of the veto channel.

The final choice for a veto to be applied to the S3 burst analysis is 
H2:LSC-PRC\_CTRL with a significance threshold 200. This yields an 
efficiency of 12\% and a dead-time of less than 0.5\%.

\section{Summary}
\label{summary}

During the S2 and S3 runs of the LIGO gravitatioanl wave burst data analyses, 
an important effort has been made in order to find efficient ways to clean 
the data from known noise sources. Several methods to isolate poor quality 
data or select effective vetoes have been studied and applied in both 
analyses. In the S2 analysis case, no good veto was found. For the 
S3 analysis a veto has been proposed, based on the channel LSC-PRC\_CTRL 
high-pass filtered at 70Hz, for the Hanford 2km detector.

\ack

The authors gratefully acknowledge the support of the United States National 
Science Foundation for the construction and operation of the LIGO Laboratory 
and the Particle Physics and Astronomy Research Council of the United Kingdom, 
the Max-Planck-Society and the State of Niedersachsen/Germany for support of 
the construction and operation of the GEO600 detector. The authors also 
gratefully acknowledge the support of the research by these agencies and by 
the Australian Research Council, the Natural Sciences and Engineering Research 
Council of Canada, the Council of Scientific and Industrial Research of India, 
the Department of Science and Technology of India, the Spanish Ministerio de 
Educacion y Ciencia, the John Simon Guggenheim Foundation, the Leverhulme 
Trust, the David and Lucile Packard Foundation, the Research Corporation, 
and the Alfred P. Sloan Foundation.

\end{document}